\documentclass{article}
\usepackage{ijcai26}
\usepackage{times}
\usepackage{soul}
\usepackage[utf8]{inputenc}
\usepackage[T1]{fontenc} 
\usepackage{url}
\usepackage{amsmath}
\usepackage{amsthm}
\usepackage{amssymb}

\usepackage[table]{xcolor} 
\usepackage{graphicx}
\usepackage{tikz}
\usetikzlibrary{positioning,fit,calc}
\usepackage{forest}
\usepackage{fontawesome5}
\usepackage{academicons}

\usepackage{tabularx}
\usepackage{booktabs} 
\usepackage{multirow}
\usepackage{makecell}
\usepackage{colortbl} 
\usepackage{array}
\usepackage{ragged2e}

\usepackage{caption}
\usepackage{subcaption}
\usepackage{cuted} 

\usepackage{algorithm}
\usepackage{algorithmic}

\usepackage[switch]{lineno} 

\usepackage[hidelinks]{hyperref}
\newcommand{\citetight}[1]{\mbox{\cite{#1}}}

\definecolor{radiantK}{HTML}{0056B3}
\definecolor{radiantP}{HTML}{20C997}
\definecolor{radiantA}{HTML}{FD7E14}
\definecolor{fusionK}{HTML}{4338CA}
\definecolor{fusionP}{HTML}{06B6D4}
\definecolor{fusionA}{HTML}{F43F5E}
\definecolor{clinicalK}{HTML}{2563EB}
\definecolor{clinicalP}{HTML}{10B981}
\definecolor{clinicalA}{HTML}{8B5CF6}
\definecolor{contrastK}{HTML}{1E293B}
\definecolor{contrastP}{HTML}{3B82F6}
\definecolor{contrastA}{HTML}{F59E0B}
\definecolor{researchK}{HTML}{0F766E}
\definecolor{researchP}{HTML}{84CC16}
\definecolor{researchA}{HTML}{EC4899}

\def\colorscheme{5} 

\ifnum\colorscheme=1
  \colorlet{mainK}{radiantK} \colorlet{mainP}{radiantP} \colorlet{mainA}{radiantA}
\fi
\ifnum\colorscheme=2
  \colorlet{mainK}{fusionK} \colorlet{mainP}{fusionP} \colorlet{mainA}{fusionA}
\fi
\ifnum\colorscheme=3
  \colorlet{mainK}{clinicalK} \colorlet{mainP}{clinicalP} \colorlet{mainA}{clinicalA}
\fi
\ifnum\colorscheme=4
  \colorlet{mainK}{contrastK} \colorlet{mainP}{contrastP} \colorlet{mainA}{contrastA}
\fi
\ifnum\colorscheme=5
  \colorlet{mainK}{researchK} \colorlet{mainP}{researchP} \colorlet{mainA}{researchA}
\fi

\newcommand{\Ktag}[1]{\cellcolor{mainK!15}\textbf{#1}}
\newcommand{\Ptag}[1]{\cellcolor{mainP!15}\textbf{#1}}
\newcommand{\Atag}[1]{\cellcolor{mainA!15}\textbf{#1}}

\newcolumntype{Y}{>{\RaggedRight\arraybackslash}X} 
\newcolumntype{S}[1]{>{\centering\arraybackslash}p{#1}} 
\newcolumntype{L}{>{\raggedright\arraybackslash}X} 

\title{Knowledge-Guided 3D CT Generation: A Conditioning-Centric Taxonomy}

\author{
Francesca Pia Panaccione
\and
Eugenio Lomurno\and
Matteo Matteucci
\affiliations
Department of Electronics, Information, and Bioengineering\\Politecnico di Milano
\emails
francescapia.panaccione@polimi.it,
eugenio.lomurno@polimi.it,
matteo.matteucci@polimi.it
}

\begin{document}

\maketitle

\begin{abstract}
Controllable generation guided by external knowledge is a key requirement in modern generative deep learning applications, enabling the synthesis of samples with explicit constraints on semantic content, structural properties, and variability. In 3D Computed Tomography (CT), such control is essential for clinical applications, including data augmentation, privacy-preserving data sharing, and the simulation of specific anatomical or pathological scenarios. While research on conditional 3D CT generation has expanded rapidly, the diversity of existing approaches makes systematic comparison 
difficult and obscures fundamental design choices.
In this survey, we propose a conditioning-centric taxonomy that organizes the literature along three orthogonal dimensions: the type of external knowledge (K), the knowledge integration paradigm (I), and the generative architecture (A). This factorization defines an explicit design space ($\mathcal{K} \times \mathcal{I} \times \mathcal{A}$) that provides a unified perspective on prior work. Using this framework, we systematize existing methods, identify dominant trends and recurring design patterns, and highlight underexplored regions of the design space that point toward promising directions for future research.
\end{abstract}

\section{Introduction}
Recent advances in deep generative modeling have enabled the synthesis of complex data across multiple domains. 
In medical imaging, synthetic data generation addresses limitations of real-world datasets such as high acquisition costs, annotation requirements, and privacy constraints~\cite{koetzier2024generating,lomurno2025federated}.
Three-dimensional (3D) medical imaging modalities represent volumetric data as spatially correlated slices, offering richer anatomical information than two-dimensional imaging~\cite{wu2025vision} while requiring generative models to maintain coherence across the entire volume~\cite{friedrich2024deep}. Among these, 3D computed tomography (CT) is increasingly adopted in clinical practice for diagnosing a wide range of conditions.

\begin{figure}[h]
    \centering
    \includegraphics[width=1.\linewidth]{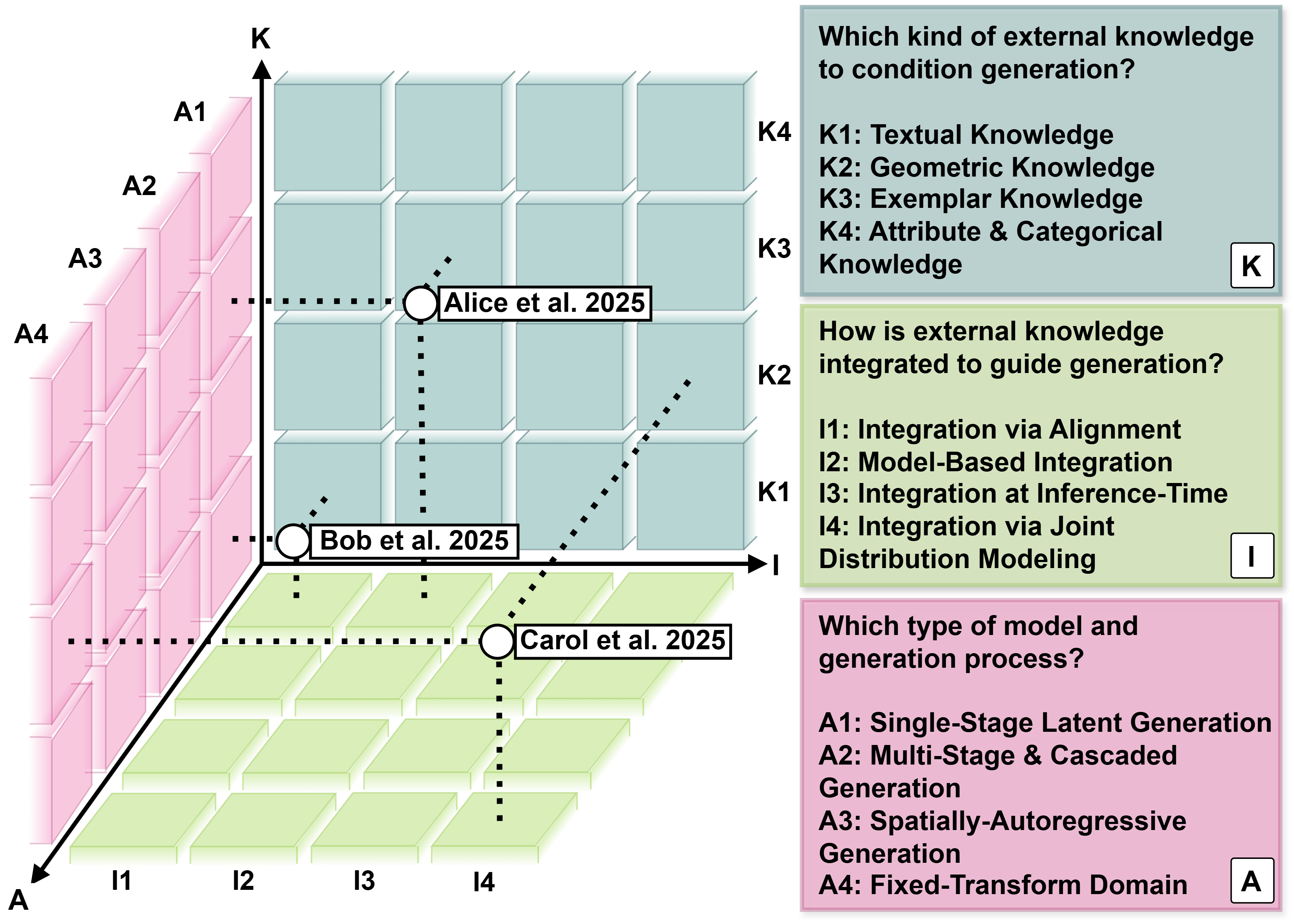}
    \caption{Conceptual representation of the proposed taxonomy as a three-dimensional design space. Each knowledge-guided 3D CT generation method can be positioned as a point in $\mathcal{K} \times \mathcal{I} \times \mathcal{A}$, defined by its choices along three orthogonal axes: external knowledge type ($\mathcal{K}$), integration paradigm ($\mathcal{I}$), and generative architecture ($\mathcal{A}$).}
    \label{fig:taxonomy_cube}
\end{figure}

Early approaches to 3D CT generation focused on unconditional modeling, learning the data distribution directly from volumetric samples, but often failed to control artifacts or ensure semantic consistency~\cite{friedrich2024deep}.
Supported by large-scale public CT datasets~\cite{hamamci2024foundation}, recent work has shifted toward \textit{knowledge-guided} generative frameworks.
These incorporate structured priors—semantic descriptors, anatomical constraints, or population attributes—to bias generation toward anatomically consistent, clinically meaningful volumes~\cite{dorjsembe2024conditional,xu2024medsyn,guo2025maisi,yoon2025cascaded}.

This rapid growth has revealed that guidance effectiveness depends critically on knowledge type and integration mechanism~\cite{hamamci2024generatect,amirrajab2025radiology,molino2025text}.
Yet the conditioning process itself remains systematically unexamined. Existing surveys organize the field by architectural families or application domains~\cite{zhou2025generative,friedrich2024deep,chen2024deep,liu20243d}, leaving trends, failure modes, and principled research directions obscured.

This survey addresses this gap by reframing the literature through a conditioning-centric perspective, examining how knowledge is represented, injected, and exploited throughout generation.
To this end we present an innovative faceted taxonomy (Figure~\ref{fig:taxonomy_cube}) that organizes methods along three orthogonal dimensions: (i) the type of external knowledge used for conditioning, (ii) how such knowledge is exploited within the conditioning paradigm, and (iii) the underlying generative process and model architecture. Together, these dimensions define a three-dimensional design space in which methods can be systematically positioned and compared.
To summarize, the contributions of this survey are as follows:
\begin{itemize}
\item \textbf{A conditioning-centric taxonomy} for 3D CT generation, enabling principled positioning of existing methods and remaining extensible to future work. 
\item \textbf{A unified design-space analysis} grounded in the interpretable $\mathcal{K} \times \mathcal{I} \times \mathcal{A}$ space, revealing dominant paradigms and underexplored regions.
\item \textbf{Clear research directions} from explicit design-space gaps, identifying actionable strategies beyond incremental refinements.
\end{itemize}
To facilitate adoption, we provide an open-source repository with 
reference implementations and an interactive classification tool at 
\url{https://github.com/eugeniolomurno/3D-CT-taxonomy}.

\section{Preliminaries}
This section formalizes knowledge-guided 3D CT generation as a constrained stochastic sampling problem over high-dimensional volumetric data.

\paragraph{Knowledge-Guided 3D CT Generation.}
Let $x \in \mathcal{X} \subset \mathbb{R}^{H \times W \times D}$ denote a 3D CT volume represented as a voxel-based grid with inherent spatial coherence, and let $k \in \mathcal{K}$ denote external information available at generation time. 
Knowledge-guided CT generation is formalized as sampling from a conditional distribution $p_\theta(x \mid k)$, with $\theta$ parameterizing the generative model and $k$ assumed exogenous to the volumetric state.
For fixed $k$, the conditional distribution retains stochastic support, admitting multiple plausible realizations that satisfy the imposed constraints. Knowledge-guided generation is therefore treated as constrained stochastic sampling, where modeling choices regulate how uncertainty is resolved under external constraints.

\paragraph{Probabilistic Conditioning.}
Conditioning can be instantiated through distinct probabilistic formulations. 
Most approaches directly parameterize $p_\theta(x \mid k)$, treating $k$ as a fixed input modulating the generative process during training.
Others model the joint distribution $p_\theta(x, k)$, where both volume and conditioning signal are co-generated, enforcing intrinsic alignment rather than treating $k$ as an external constraint. A third paradigm modifies the sampling process at inference time: the model parameters remain fixed, but the sampling trajectory is dynamically adjusted to favor configurations consistent with $k$. These formulations correspond to different probabilistic objectives and encode different assumptions on how constraints interact with generative uncertainty.

\paragraph{Distribution Decomposition in Volumetric Generation.}
The dimensionality and spatial structure of $\mathcal{X}$ require decomposing the target distribution into tractable components. Abstractly, volumetric synthesis proceeds by selecting a decomposition strategy for $p(x \mid k)$: through global stochastic processes acting on the full volume, through explicit spatial factorizations $p(x \mid k) = \prod_{n} p(x_n \mid x_{<n}, k)$ as in spatial-autoregressive schemes, through hierarchical multi-scale decompositions generating coarse-to-fine structure, or through deterministic invertible transformations $x = T^{-1}(y)$ with $y \sim p(y \mid k)$ that redistribute spatial dependencies into structured representation spaces. 
In practice, these decompositions are often implemented in learned latent spaces via an encoder--decoder pair $(E, D)$, yielding the generative objective $p_\theta(z \mid k)$ with $z = E(x)$ and $x \approx D(z)$ at reduced computational cost.

\section{Taxonomy}\label{sec:taxonomy}

This section introduces the proposed taxonomy for organizing 3D CT knowledge-guided approaches. In contrast to prior surveys, which predominantly organize the literature by architectural families, imaging modalities, or application domains (Table~\ref{tab:related-surveys})~\cite{khader2023denoising,friedrich2024deep,chen2024deep,liu20243d,zhou2025generative}, our formulation explicitly disentangles the conditional generative process in 3D CT synthesis.
Rather than ranking methods or prescribing optimal designs, the taxonomy provides a descriptive conceptual framework for organizing and interpreting the existing literature.

\begin{table}[t]
\centering
\scriptsize
\setlength{\tabcolsep}{1.7pt}
\renewcommand{\arraystretch}{1.0}
\begin{tabular}{@{}llllc@{}}
\toprule
\textbf{Survey} & \textbf{Organization} & \textbf{Breadth} & \textbf{Qnt. Analysis} & \textbf{Design Space} \\
\midrule
\cite{khader2023denoising} & Architecture & Med-3D & Performance & -- \\
\cite{friedrich2024deep} & Modality & Med-3D & -- & -- \\
\cite{liu20243d} & Application & Brain/Heart & Performance & -- \\
\cite{chen2024deep} & Architecture & Gen-3D & -- & -- \\
\cite{zhou2025generative} & Application & Med-All & Performance & -- \\
\midrule
\rowcolor{gray!12}
\textbf{Ours} & \textbf{Conditioning} & \textbf{3D CT} & \textbf{Distributional} & \checkmark \\
\bottomrule
\end{tabular}
\caption{Comparison with related surveys. This work introduces the first conditioning-centric taxonomy with explicit method positioning a structured design space ($\mathcal{K} \times \mathcal{I} \times \mathcal{A}$).}
\label{tab:related-surveys}
\end{table}

\subsection{Taxonomy Structure}\label{subsec:tax_structure}
The taxonomy is structured as a three-dimensional design space $\mathcal{K} \times \mathcal{I} \times \mathcal{A}$ that factorizes knowledge-guided 3D CT generation methods along three independent axes (Figure~\ref{fig:taxonomy_cube}).
Each axis captures a distinct structural aspect of model design: the type of external knowledge involved (Axis~$\mathcal{K}$), the mechanism by which such knowledge interacts with the generative process (Axis~$\mathcal{I}$), and the strategy adopted for volumetric synthesis (Axis~$\mathcal{A}$).
Within this structure, each method is represented as a tuple $(k,i,a)$, where $k \subseteq \mathcal{K}$, $i \subseteq \mathcal{I}$, and $a \subseteq \mathcal{A}$. All components are non-empty, and multiple categories per axis are permitted to account for hybrid approaches. 
This explicit factorization enables systematic comparison across heterogeneous methods, making recurrent patterns as well as sparsely explored regions of the design space directly observable.
The categories defining each axis are detailed in the following subsections, with representative instantiations summarized in Tables~\ref{tab:taxonomy-K}, \ref{tab:taxonomy-I}, and \ref{tab:taxonomy-A}.

\begin{table}[t]
\centering
\footnotesize
\renewcommand{\arraystretch}{1.15}
\setlength{\tabcolsep}{5.5pt}
\begin{tabularx}{\columnwidth}{@{}>{\raggedright\arraybackslash}p{2.8cm}X@{}}
\toprule
\rowcolor{mainK!18}
\multicolumn{2}{@{}l}{\textbf{Axis $\mathcal{K}$: External Knowledge}}\\
\midrule
\textbf{Category} & \textbf{Representative Instantiations} \\
\midrule
\rowcolor{mainK!4}
\textbf{K1} - Textual Knowledge &
Radiology reports, free-text clinical descriptions, open-vocabulary prompts, unstructured findings. \\
\textbf{K2} - Geometric Knowledge &
Organ segmentation masks, anatomical layouts, bounding boxes, landmark coordinates, sparse structural maps. \\
\rowcolor{mainK!4}
\textbf{K3} - Exemplar Knowledge &
Reference volumes from complementary modalities (e.g., MRI, PET, CBCT), patient-specific priors, atlas templates. \\
\textbf{K4} - Attribute \& Categorical Knowledge &
Patient demographics (e.g., age, sex, BMI), diagnostic class labels, acquisition metadata, structured clinical attributes. \\
\bottomrule
\end{tabularx}
\caption{Taxonomy of Axis $\mathcal{K}$ (External Knowledge). Categories represent different types of information used to condition 3D CT generation, ranging from unstructured text to dense volumetric priors.}
\label{tab:taxonomy-K}
\end{table}

\subsection{Axis $\mathcal{K}$: External Knowledge}\label{subsec:axis_k}
This axis characterizes the form of external knowledge used to condition generation. 
Consistently with the problem formulation in Section~2, external knowledge $k \in \mathcal{K}$ is defined as information exogenous to the CT volume. 
Different instantiations of $k$ vary in modality and in the degree of constraint they impose on generation. Based on these differences, we identify four broad categories, described below and illustrated with representative examples in Table~\ref{tab:taxonomy-K}: 
\paragraph{K1: Textual Knowledge.}K1 corresponds to unstructured semantic information encoded in linguistic form and mapped to latent representations. 
It provides high-level semantic conditioning without explicitly enforcing structural constraints, such that anatomical localization and organization are expected to emerge implicitly through the generative dynamics.

\paragraph{K2: Geometric Knowledge.}K2 encodes constraints on anatomical structure and spatial organization, either through voxel-aligned representations or abstract geometric descriptors.
While these priors enforce structural plausibility, they abstract away non-geometric variation, limiting the modeling of physiological patterns not directly captured by geometry.

\paragraph{K3: Exemplar Knowledge.}K3 comprises dense volumetric priors from reference instances. By 
providing voxel-aligned structural and appearance information, exemplar knowledge anchors generation to specific anatomical realizations, constraining synthesis at the instance level.

\paragraph{K4: Attribute \& Categorical Knowledge.}K4 encodes non-spatial descriptive variables associated with the target volume, enabling global modulation of the data distribution. This conditioning captures population-level regularities linked to phenotypic or acquisition-related factors without prescribing local spatial structure.

\begin{table}[t]
\centering
\footnotesize
\renewcommand{\arraystretch}{1.2}
\setlength{\tabcolsep}{6pt}
\begin{tabularx}{\columnwidth}{@{}>{\raggedright\arraybackslash}p{2.8cm}X@{}}
\toprule
\rowcolor{mainP!18}
\multicolumn{2}{@{}l}{\textbf{Axis $\mathcal{I}$: Knowledge Integration}}\\
\midrule
\textbf{Category} & \textbf{Representative Instantiations} \\
\midrule
\rowcolor{mainP!4}
\textbf{I1} - Integration via Alignment  &
Pre-generative embedding alignment (e.g., CLIP, BioViL), dual-encoder architectures, contrastive representation learning. \\
\textbf{I2} - Model-Based Integration &
Cross-attention layers, channel-wise concatenation, feature-wise modulation (FiLM, AdaGN), ControlNet adapters. \\
\rowcolor{mainP!4}
\textbf{I3} - Integration at Inference-Time &
Classifier-free guidance, energy-based Guidance, trajectory smoothing, compositional guidance. \\
\textbf{I4} - Integration via
Joint distribution modeling &
joint probability factorization $p(x,k)$, unified diffusion over concatenated modalities (e.g., image + mask). \\
\bottomrule
\end{tabularx}
\caption{Taxonomy of Axis $\mathcal{I}$ (Knowledge Integration). Categories distinguish the stage and mechanism through which external knowledge influences the generative process, from pre-generative alignment to joint distribution modeling.}
\label{tab:taxonomy-I}
\end{table}

\subsection{Axis $\mathcal{I}$: Knowledge Integration}\label{subsec:axis_i}
This axis characterizes how external knowledge influences the generative process, independently of its instantiation (Axis $\mathcal{K}$) or the architectural backbone (Axis $\mathcal{A}$). We identify four paradigms (Table~\ref{tab:taxonomy-I}).
Throughout this section, the generative target is denoted by $x$, representing either volumetric data or its latent representation.

\paragraph{I1: Integration via Alignment.}
I1 paradigms enforce conditioning by establishing a correspondence between external knowledge $k$ and $x$ within a shared representation space prior to generation. Both modalities are mapped to compatible representations aligned under a common similarity or consistency criterion. Conditioning is thus imposed implicitly at the representation level, with knowledge influencing generation through alignment rather than direct intervention in the model’s internal dynamics.

\paragraph{I2: Model-Based Integration.}
In I2 paradigm, external knowledge $k$ is incorporated directly into the generator architecture to condition the generative dynamics, corresponding to learning a conditional predictor (e.g., $\epsilon_\theta(x_t, t \mid k)$) in which knowledge modulates the model’s internal computations. The integration mechanism depends on the structural properties of $k$: non-spatial or abstract signals induce global modulation, whereas spatially aligned knowledge enable localized conditioning that preserves correspondence. Conditioning is embedded within the learned model and consistently guides generation.

\paragraph{I3: Integration at Inference-Time.}
I3 paradigms apply conditioning exclusively at inference time, leaving the learned generative distribution unchanged. In diffusion-based models, a canonical instantiation is classifier-free guidance, which linearly combines unconditional and conditional predictions at each diffusion step $t$ as $\tilde{\epsilon}(x_t \mid k) = \epsilon_\theta(x_t, t) + s \cdot (\tilde{\epsilon}\theta(x_t, t \mid k) - \epsilon\theta(x_t, t))$, where $s$ controls the fidelity–diversity trade-off during reverse diffusion.

\paragraph{I4: Integration via Joint Distribution Modeling.}
Unlike prior paradigms where $k$ is fixed, I4 methods model the joint distribution $p_\theta(x, k)$, generating knowledge and target variables simultaneously. This enforces intrinsic coupling between $x$ and $k$, enabling jointly emerging structure and appearance. While more computationally demanding, I4 approaches remove conditioning asymmetry and capture bidirectional correlations.

\subsection{Axis $\mathcal{A}$: Generative Architecture}\label{subsec:axis_a}
This axis characterizes the architectural strategy for volumetric synthesis, independently of knowledge representation ($\mathcal{K}$) or integration ($\mathcal{I}$). It describes how the high-dimensional generation problem is decomposed into tractable subproblems. We identify four strategies based on scope and representation space (Table~\ref{tab:taxonomy-A}).

\begin{table}[t]
\centering
\footnotesize
\renewcommand{\arraystretch}{1.2}
\setlength{\tabcolsep}{6pt}
\begin{tabularx}{\columnwidth}{@{}>{\raggedright\arraybackslash}p{2.8cm}X@{}}
\toprule
\rowcolor{mainA!18}
\multicolumn{2}{@{}l}{\textbf{Axis $\mathcal{A}$: Generative Architecture}}\\
\midrule
\textbf{Category} & \textbf{Representative Instantiations} \\
\midrule
\rowcolor{mainA!4}
\textbf{A1} - Single-Stage Latent Generation &
Holistic generation in learned latent spaces, 3D-VQGAN backbones, one-shot synthesis. \\
\textbf{A2} - Multi-Stage \& Cascaded Generation &
Coarse-to-fine synthesis, super-resolution cascades, hierarchical residual modeling. \\
\rowcolor{mainA!4}
\textbf{A3} - Spatially-Autoregressive Generation &
Slice-wise autoregression, video-like generation (z-axis evolution), temporal transformers, spatial decomposition. \\
\textbf{A4} - Fixed-Transform Domain &
Wavelet-domain diffusion, spectral diffusion, generation in deterministic invertible spaces. \\
\bottomrule
\end{tabularx}
\caption{Taxonomy of Axis $\mathcal{A}$ (Generative Architecture). Categories characterize the structural strategy for volumetric synthesis, distinguishing methods by their procedural decomposition and representation space.}
\label{tab:taxonomy-A}
\end{table}

\paragraph{A1: Single-Stage Latent Generation.}
A1 strategies model volumetric synthesis holistically by generating either the full volume $x$ or a learned latent embedding from which $x$ is decoded. Generation is performed in a single stage, without explicit hierarchical or spatial decomposition. This formulation favors computational efficiency and global coherence, relying on the expressive capacity of the latent representation to capture structural consistency.

\paragraph{A2: Multi-Stage \& Cascaded Generation.}
A2 strategies decompose volumetric synthesis into a sequence of conditional stages, typically operating at increasing resolutions. Generation follows a coarse-to-fine hierarchy, with $x^{(1)} \sim p_{\theta_1}(x^{(1)})$ and $x^{(2)} \sim p_{\theta_2}(x^{(2)} \mid x^{(1)})$, where the initial stage captures low-resolution global structure and subsequent stages progressively refine local details.

\paragraph{A3: Spatially-Autoregressive Generation.}
A3 methods factorize the generation along a spatial axis (typically the axial depth $D$), treating the 3D volume as a sequence of 2D slices or sub-volumes.
The process follows an ordered dependency, where the generation of the $n$-th slice is conditioned on the previously synthesized context ($x_{<n}$). 
This formulation models inter-slice dependencies through sequential factorization.

\paragraph{A4: Fixed-Transform Domain Generation.}
A4 strategies perform synthesis in a deterministic representation space defined by a fixed, invertible transform $T$. 
Generation is carried out over transformed coefficients $y \sim p_\theta(y \mid k)$, with reconstruction via $x = T^{-1}(y)$. 
These approaches reduce dimensionality while preserving exact invertibility, at the expense of flexibility in representation learning.

\begin{table*}[!t]
\centering
\scriptsize
\setlength{\tabcolsep}{2.5pt}
\renewcommand{\arraystretch}{1.1}
\begin{tabularx}{\textwidth}{
>{\raggedright\arraybackslash}p{2.4cm}
S{0.8cm}
S{0.6cm}
S{0.5cm}
>{\raggedright\arraybackslash}p{2.2cm}
>{\raggedright\arraybackslash}p{3.2cm}
>{\raggedright\arraybackslash}p{2.5cm}
>{\raggedright\arraybackslash}p{3.4cm}
c
}
\toprule
\textbf{Method} & \textbf{K} & \textbf{I} & \textbf{A} &
\textbf{Conditioning Source} &
\textbf{Conditioning Mechanism} &
\textbf{Generation Strategy} &
\textbf{Dataset (Resolution)} &
\textbf{Code} \\
\midrule

MedGen3D~\cite{han2023medgen3d} &
\Ktag{K2} & \Ptag{I4} & \Atag{A3} &
Voxel-aligned Semantic Maps &
Joint mask--image diffusion (MC-DPM) &
Autoregressive slice-wise &
SegTHOR (96$\times$320$^2$)&
-- \\

GenerateCT~\cite{hamamci2024generatect} &
\Ktag{K1} & \Ptag{I1, I2} & \Atag{A2} &
Free-form Text Prompts &
Cross-attention, CFG &
Cascaded low-res$\rightarrow$sup-res &
CT-RATE (512$^2\times$201) &
\href{https://github.com/ibrahimethemhamamci/GenerateCT}{\faGithub} \\

MedSyn~\cite{xu2024medsyn} &
\Ktag{K1, K2} & \Ptag{I4} & \Atag{A2} &
Textual Description + Anatomical Semantic Layouts &
Joint Diffusion
&
Multi-Staged low-res$\rightarrow$sup-res &
Private Lung Dataset (256$^3$) &
\href{https://github.com/batmanlab/MedSyn}{\faGithub} \\

GEM-3D~\cite{zhu2024generative} &
\Ktag{K2, K3} & \Ptag{I2} & \Atag{A3} &
Anatomical Masks + Reference slice &
Latent Concatenation &
Sequential Window-based &
AbdomenCT-1K (512$^2\times$Z, Z variable) &
\href{https://github.com/HKU-MedAI/GEM-3D}{\faGithub} \\

CM3dLDM~\citetight{tapp2024mr} &
\Ktag{K3} & \Ptag{I2} & \Atag{A1} &
Cross-modal Volume (MRI) &
Frozen Encoder Injection &
Single-stage Patch-based Diffusion &
Private MR-CT Head Dataset; SynthRad (224$^3$) &
\href{https://github.com/AustinTapp/CM3dLDM}{\faGithub} \\

cWDM~\cite{friedrich2024cwdm} &
\Ktag{K3} & \Ptag{I2} & \Atag{A4} &
Cross-modal Volume (MRI) &
Wavelet-domain Feature Modulation &
Fixed-transform Wavelet Diffusion &
BraTS 2024 ($128^3$) &
\href{https://github.com/pfriedri/cwdm}{\faGithub} \\

DiffTumor~\cite{chen2024towards} &
\Ktag{K2, K3} & \Ptag{I2} & \Atag{A1} &
Masks + Healthy CT Volume &
Latent Patch Concatenation &
Single-stage Latent Diffusion (Inpainting) &
LiTS; MSD; KiTS; AbdAtlas-8K; Hopkins (96$^3$) &
\href{https://github.com/MrGiovanni/DiffTumor}{\faGithub} \\

MC-IDDDPM~\cite{pan2024synthetic} &
\Ktag{K3} & \Ptag{I2} & \Atag{A4} &
Cross-modal Volume (MRI) &
Feature Concatenation &
Single-stage Diffusion &
Private Brain and Prostate Datasets MRI-CT (192$^2\times$96) &
-- \\

LN-DDPM~\cite{yu2024ct} &
\Ktag{K2} & \Ptag{I2} & \Atag{A1} &
Lymph Node + Organ Masks &
Spatial Concatenation &
Single-stage patch-based diffusion &
Private Colorectal Dataset; ABD-LN; (128$^3$) &
-- \\

Med-LVDM~\cite{kui2025med} &
\Ktag{K3} & \Ptag{I2} & \Atag{A1} &
Cross-modal Volume
(MRI) &
Latent concatenation  &
Single-stage Latent Diffusion &
Pelvis MR--CT (256$^2\times$Z) &
\href{https://github.com/mirthAI/Med3DVLM}{\faGithub} \\

Cascaded-3D~\cite{yoon2025cascaded} &
\Ktag{K4} & \Ptag{I2} & \Atag{A2} &
Demographics &
AdaGN &
Cascaded coarse$\rightarrow$ sup-res &
AutoPET (224$\times$224$\times$384) &
-- \\

Surf2CT~\cite{yoon2025surf2ct} &
\Ktag{K2,K4} & \Ptag{I2} & \Atag{A2} &
Skin Surface + Demographics &
Conditional flow matching &
Cascaded coarse$\rightarrow$super-res &
Private Torso CT Dataset; AutoPET (224$\times$224$\times$352) &
-- \\

CTFlow~\cite{wang2025ctflow} &
\Ktag{K1} & \Ptag{I2} & \Atag{A3} &
Clinical Reports &
Cross-attention &
Slice-as-video Flow Matching &
CT-RATE (256$^2\times$Z, Z variable) &
-- \\

Report2CT~\cite{amirrajab2025radiology} &
\Ktag{K1} & \Ptag{I2, I3} & \Atag{A1} &
Clinical Reports &
Multi-encoder Cross-Attention, CFG &
Single-stage Latent Diffusion &
CT-RATE (480$^2\times$256) &
\href{https://github.com/sinaamirrajab/report2ct}{\faGithub} \\

Text-to-CT~\cite{molino2025text} &
\Ktag{K1} & \Ptag{I1, I2} & \Atag{A1} &
Radiology text &
3D CLIP pretraining, Cross-Attention &
Single-stage Latent Diffusion &
CT-RATE (512$^2\times$128) &
\href{https://github.com/danielemolino/Text2CT}{\faGithub} \\

Text2CT~\cite{guo2025text2ct} &
\Ktag{K1} & \Ptag{I2, I3} & \Atag{A1} &
Free-text descriptions &
Cross-attention, CFG &
Single-stage Latent Diffusion &
CT-RATE; RadChestCT (512$^2\times$192) &
-- \\

LAND~\cite{oliveras2025land} &
\Ktag{K2} & \Ptag{I2} & \Atag{A1} &
Lung + Nodule Segmentation Masks &
Latent Concatenation, Cross Attention &
Single-stage Latent Diffusion &
LIDC-IDRI; NLST (256$^3$) &
-- \\

MedLoRD~\cite{seyfarth2025medlord} &
\Ktag{K2} & \Ptag{I2} & \Atag{A1} &
Anatomical Segmentation Masks &
ControlNet &
Single-stage Latent Diffusion &
Private Coronary Dataset; LUNA16 (512$^2\times$256) &
\href{https://github.com/Cardio-AI/medlord}{\faGithub} \\

MAISI~\cite{guo2025maisi} &
\Ktag{K2} & \Ptag{I2} & \Atag{A1} &
Multi-organ Segmentation Maps &
ControlNet &
Single-stage Latent Diffusion &
Curated multi-source, multi-organ CT dataset (512$^2\times$768) &
\href{https://github.com/Project-MONAI/tutorials/blob/main/generation/maisi}{\faGithub} \\

MAISI-v2~\cite{zhao2025maisi} &
\Ktag{K2} & \Ptag{I2} & \Atag{A1} &
Multi-organ Segmentation Maps &
ControlNet, RCL &
Single-stage latent &
Curated multi-source multi-organ CT dataset (512$^2\times$768)  &
\href{https://github.com/NVIDIA-Medtech/NV-Generate-CTMR/tree/main}{\faGithub} \\

NodMAISI~\cite{tushar2025nodmaisi} &
\Ktag{K2} & \Ptag{I2} & \Atag{A1} &
Nodule-specific Semantic Masks &
ControlNet &
Single-stage Latent Diffusion &
Curated multi-source public lung CT datasets (512$^2\times$768) &
\href{https://github.com/fitushar/NoMAISI}{\faGithub} \\

TRACE~\cite{shao2025trace} &
\Ktag{K1, K2} & \Ptag{I2, I3} & \Atag{A3} &
Text + Multi-modal Anatomical Masks &
Multi-modal Mask Concatenation &
Autoregressive Slice-pair (Video-like) &
CT-RATE (256$^2\times$Z, Z variable) &
\href{https://github.com/VinyehShaw/TRACE}{\faGithub} \\

Lung-DDPM~\cite{jiang2025lung} &
\Ktag{K2, K3} & \Ptag{I2} & \Atag{A1} &
Semantic Layouts + Reference CT &
Layout concatenation + AAS &
Single-stage Layout-guided &
Private Lung Dataset; LIDC-IDRI (128$^3$) &
\href{https://github.com/Manem-Lab/Lung-DDPM}{\faGithub} \\

3DLDM~\cite{mahdi20253d} &
\Ktag{K3} & \Ptag{I2} & \Atag{A1} &
Cross-modal Volume (MRI) &
Latent Feature Concatenation &
Single-stage Latent Diffusion &
SynthRAD2023 (96$^2\times$256) &
-- \\

3D-WLDM~\cite{zheng20253d} &
\Ktag{K3} & \Ptag{I2} & \Atag{A4} &
Cross-modal Volume (MRI) &
Latent Wavelet Concatenation &
Fixed-transform Latent Diffusion &
Private PET/MR–PET/CT Datasets (128$^3$) &
-- \\

LabelG~\cite{wang2025labelg} &
\Ktag{K2} & \Ptag{I4} & \Atag{A1} &
Semantic Segmentation Masks &
Joint Latent-space Modeling &
Single-stage Latent Diffusion&
Private Abdominal CT dataset, AbdomenCT; SegTHOR; MSD10-Colon; (256$\times$256$\times$128) &
-- \\

\bottomrule
\end{tabularx}
\caption{Comprehensive overview of knowledge-guided 3D CT generation methods organized under the proposed taxonomy. For each method, we report the classification along axes K, I, and A, the conditioning source, the conditioning mechanism, the generation strategy, the dataset used with volumetric resolution (as reported by authors when available), and code availability. Methods using multiple categories within a single axis are listed with all applicable labels. Note: CFG = Classifier-Free Guidance; AdaGN = Adaptive Group Normalization; RCL = Region-specific Contrastive Loss; AAS = Anatomically-Aware Sampling; MC-DPM = Multi-Condition Denoising Probabilistic Model. \faGithub\,indicates publicly available code repository.}
\label{tab:kg-ct}
\end{table*}

\begin{figure*}[t]
    \centering
    \includegraphics[width=1\linewidth]{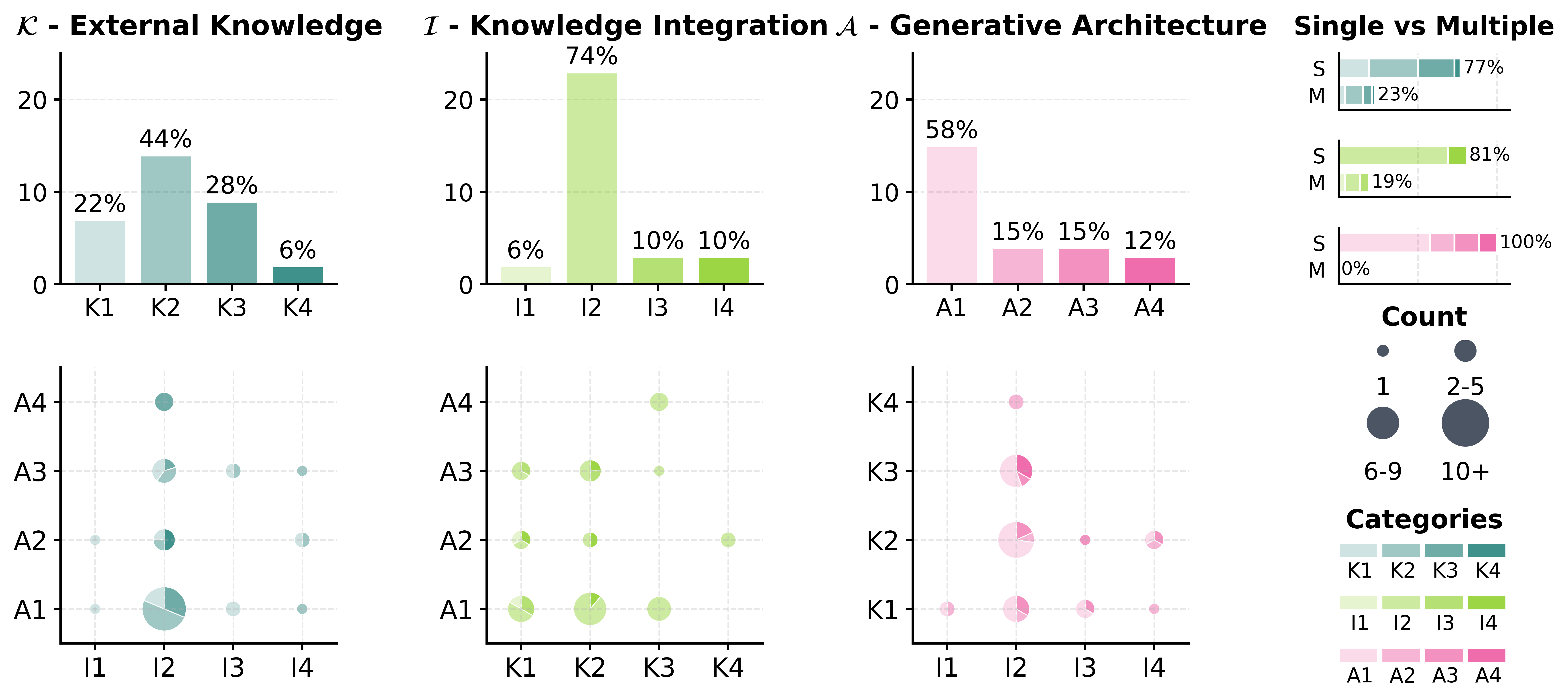}
    \caption{Statistical distribution of methods across the taxonomy. Top row: frequency of each category in axes $\mathcal{K}$, $\mathcal{I}$, and $\mathcal{A}$, plus single versus multiple category usage per axis. Bottom row: pairwise interactions between axes, where node size indicates method count and pie slices show the third axis's distribution (methods using multiple categories appear in multiple counts).}
    \label{fig:summary}
\end{figure*}


\section{Literature Trends}\label{sec:literature}
This section applies the proposed taxonomy (\(\mathcal{K}\times\mathcal{I}\times\mathcal{A}\)) to organize and analyze the literature on knowledge-guided 3D CT generation. Table~\ref{tab:kg-ct} provides a comprehensive overview of all reviewed methods, while Figure~\ref{fig:summary} quantifies the distribution of approaches across the three axes. The analysis is structured thematically, identifying dominant trends, recurring patterns, and underexplored regions in the proposed design space.

\paragraph{External Knowledge Trends.}
Figure~\ref{fig:summary} (top-left) shows the distribution of methods across knowledge categories.
\emph{Geometric knowledge} (K2) is the most prevalent, accounting for 44\% of methods, instantiated primarily through organ segmentation masks. Representative approaches include MAISI, MAISI-v2, NodMAISI, LAND, and MedLoRD~\cite{guo2025maisi,zhao2025maisi,tushar2025nodmaisi,oliveras2025land,seyfarth2025medlord}.

\emph{Exemplar conditioning} (K3) accounts for 28\% of existing methods and is predominantly used in cross-modal translation settings (e.g., MRI$\rightarrow$CT). Representative approaches such as 3DLDM, 3D-WLDM, and Med-LVDM leverage source latents as structural exemplars, enabling strong alignment between input and output. Hybrid K2+K3 strategies, including DiffTumor and Lung-DDPM, demonstrate lesion-aware or layout-guided synthesis with high structural fidelity~\cite{mahdi20253d,zheng20253d,kui2025med,chen2024towards,jiang2025lung}.

\emph{Text-based conditioning} (K1) accounts for 22\% of existing methods. Representative approaches include GenerateCT, Text-to-CT, Report2CT, and Text2CT, which adopt medical-specific encoders or task-adapted language models to obtain richer semantic representations~\cite{hamamci2024generatect,molino2025text,amirrajab2025radiology,guo2025text2ct}.
\emph{Distributional knowledge} (K4) accounts for 6\% of existing methods. Representative examples include Cascaded-3D and Surf2CT, as summarized in Table~\ref{tab:kg-ct}. Non-spatial descriptive variables such as demographics can be obtained without additional encoders~\cite{yoon2025cascaded,yoon2025surf2ct}.

\paragraph{Knowledge Integration Trends.}
The distribution across integration paradigms (Figure~\ref{fig:summary}, top-center) reveals strong concentration. 
\emph{Model-based integration} (I2) accounts for 74\% of methods, incorporating external knowledge directly into the generator architecture. Representative approaches employ mechanisms such as cross-attention or feature concatenation to inject conditioning signals, enabling compatibility with all knowledge types (K1–K4) and architectural strategies. This flexibility has made I2 the most prevalent strategy for 3D CT generation~\cite{hamamci2024generatect,guo2025text2ct,oliveras2025land,mahdi20253d,kui2025med}.
\emph{Inference-time integration} (I3) and \emph{joint modeling} (I4) each comprise 10\%. A strong coupling between K1 and I3 is observed (e.g., GenerateCT, Report2CT~\cite{amirrajab2025radiology}, Text2CT~\cite{guo2025text2ct}, TRACE~\cite{shao2025trace}), while geometric (K2) and exemplar (K3) approaches do not adopt this paradigm. 
\emph{Pre-generative alignment} (I1) accounts for 6\% of existing methods and is currently confined to text-conditioned generation (K1), utilizing transformer-based tokenization and contrastive objectives~\cite{hamamci2024generatect,molino2025text}.

\paragraph{Architectural Trends.}
Figure~\ref{fig:summary} (top-center) shows that \emph{single-stage latent generation} (A1) accounts for 56\% of existing methods. Representative approaches such as MAISI and MAISI-v2 leverage VQ-VAE latent spaces, while LAND, MedLoRD, and DiffTumor adopt similar strategies~\cite{guo2025maisi,zhao2025maisi,oliveras2025land,seyfarth2025medlord,chen2024towards}.
\emph{Cascaded pipelines} (A2) decompose synthesis into coarse-to-fine stages and account for 15\% of methods. Representative examples include GenerateCT, MedSyn, and Cascaded-3D~\cite{hamamci2024generatect,xu2024medsyn,yoon2025cascaded}.
\emph{Autoregressive architectures} (A3) represent 15\% of methods and factorize generation along the axial dimension. By treating volumes as ordered slice sequences, approaches such as TRACE and CTFlow model inter-slice dependencies~\cite{shao2025trace,wang2025ctflow}.
\emph{Fixed-representation generation} (A4) accounts for 12\% of methods. Representative approaches include cWDM and 3D-WLDM, which operate in wavelet space~\cite{friedrich2024cwdm,zheng20253d}.

\paragraph{Cross-Dimensional Patterns.}
The scatter plots in Figure~\ref{fig:summary} (bottom row) reveal strong dependencies between knowledge type, integration strategy, and architectural design, highlighting both consolidated practices and systematic gaps in the design space.
A clear asymmetry emerges along the knowledge–integration axis (Figure~\ref{fig:summary}, K–I). Textual conditioning (K1) consistently co-occurs with inference-time guidance (I3), while geometric (K2), exemplar (K3), and distributional (K4) knowledge rely almost exclusively on direct architectural integration (I2).
Architectural preferences further reveal distinct patterns (Figure~\ref{fig:summary}, K–A and I–A). Geometric conditioning (K2) concentrates in single-stage latent generation (A1), forming the most frequent configuration. Exemplar-based methods (K3) distribute across latent (A1) and fixed-representation architectures (A4). Textual conditioning (K1) spans latent (A1) and autoregressive designs (A3), whereas distributional knowledge (K4) appears exclusively in cascaded pipelines (A2). Model-based integration (I2) within single-stage latent models (A1) dominates the integration–architecture space.
Taken together, these trends converge on the triplet (K2, I2, A1)—geometric masks with in-process modulation in single-stage latent diffusion—as the prevailing paradigm, exemplified by MAISI, MAISI-v2, NodMAISI, LAND, and MedLoRD~\cite{guo2025maisi,zhao2025maisi,tushar2025nodmaisi,oliveras2025land,seyfarth2025medlord}. Several regions of the design space remain comparatively underrepresented.

\section{Design Space Interpretation}\label{sec:interpretation}
The concentration around geometric conditioning (K2) with in-process modulation (I2) in single-stage latent diffusion (A1) reflects pragmatic convergence rather than fundamental optimality. Automated segmentation tools made voxel-aligned priors widely available, while I2's architectural flexibility—enabling cross-attention, concatenation, or adaptive modulation across all knowledge types—made it universally applicable without structural constraints. Geometric masks resolve spatial uncertainty directly, enforcing anatomical correctness while constraining appearance variability. This established geometry as structural backbone, effective for targeted synthesis but limiting semantic diversity and population-aware generation.
This convergence reveals three fundamental design patterns. 

\emph{First}, textual conditioning (K1) pairs exclusively with classifier-free guidance (I3)~\cite{hamamci2024generatect,guo2025text2ct,shao2025trace} because linguistic descriptions exhibit irreducible semantic ambiguity: for instance, "enlarged liver" maps to diverse configurations requiring inference-time balancing between fidelity and diversity. Geometric conditioning eliminates this need by resolving spatial constraints through voxel-aligned masks, explaining why text enables semantic specification without spatial grounding while geometry ensures precision without appearance flexibility.

\emph{Second}, pre-generative alignment (I1) remains confined to text-image pairs due to established vision-language frameworks (CLIP, contrastive learning), leaving geometric-demographic (K2–K4) or cross-modal exemplar (K3–K3) alignment unexplored despite potential for population-conditioned scaffolds or registration-free correspondence. 

\emph{Third}, architectural decompositions encode distinct trade-offs: single-stage latent models (A1) depend critically on autoencoder fidelity, manifesting compression artifacts as slice discontinuities; cascaded pipelines (A2) propagate low-resolution errors through refinement stages; autoregressive methods (A3) accumulate sequential prediction drift; fixed-transform approaches (A4) avoid learned artifacts but cluster in MRI$\rightarrow$CT translation where exemplar constraints (K3) stabilize wavelet-domain synthesis, is currently inapplicable to abstract conditioning (K1, K2) lacking inherent spatial structure.

Notably, some design space gaps reflect mechanistic constraints rather than oversight. Demographic attributes (K4) are structured and unambiguous (age: 65, sex: female), lacking the variability that motivates inference-time guidance (I3) for textual descriptions—direct feature modulation suffices, rendering K4-I3 combinations redundant. Similarly, fixed-transform generation (A4) with textual conditioning (K1) remains absent because linguistic semantics lack the geometric regularity required for deterministic transform domains.

\section{Open Research Directions}\label{sec:directions}
\paragraph{Untapped Demographic Potential.}
Despite convergence on (K2, I2, A1), demographic conditioning (K4) accounts for only 6\% of methods yet offers unique advantages: attributes are acquired without segmentation pipelines or learned encoders, enabling direct population-level modeling of age-related atrophy, sex-specific morphology, or pathology prevalence at minimal annotation cost. Combining K4 with geometric scaffolds (K2) through pre-generative alignment (I1) could establish population-conditioned anatomical priors, while K4 integration via adaptive normalization (I2) could modulate global appearance without spatial constraints.

\paragraph{Multimodal Fusion Strategies.}
Multimodal integration beyond geometry remains nascent. Text-exemplar combinations (K1+K3) could ground semantic descriptions in structural priors, mitigating spatial ambiguity through voxel-aligned constraints. Text-demographic pairing (K1+K4) could enable population-specific semantic generation, addressing limited spatial grounding in pure linguistic conditioning. More fundamentally, joint modeling (I4) warrants extension beyond paired image-mask generation: CT-MRI co-generation could enforce intrinsic structural alignment without explicit registration, while volume-text co-generation could prevent semantic drift inherent in conditional formulations. Combining I4 with hierarchical decompositions (A2) or fixed-transform domains (A4) could enable staged co-generation while controlling complexity.

\paragraph{Architectural Diversification.}
Architectural exploration reveals untapped potential. Fixed-transform strategies (A4) concentrate in exemplar-driven translation but could extend to coarse geometric conditioning (K2), leveraging wavelet or spectral representations to reduce memory requirements while preserving anatomical consistency without learned compression. Conversely, identifying minimal sufficient geometric priors—treating masks as structural scaffolds while secondary sources (K1, K3, K4) modulate appearance—could substantially reduce annotation overhead. 
Hybrid integration strategies combining alignment (I1) with model-based conditioning (I2), or I2 with inference-time guidance (I3), remain unexplored despite complementary strengths: I1 for representation coherence, I2 for spatial modulation, I3 for generation-time controllability.

\section{Limitations and Conclusions}\label{sec:conclusion}
This survey introduces the first conditioning-centric taxonomy for knowledge-guided 3D CT generation, organizing methods along orthogonal axes—knowledge type (K), integration paradigm (I), and generative architecture (A)—to enable systematic positioning within an interpretable design space. The taxonomy provides a framework for comparing conditioning strategies across heterogeneous approaches, though reliable performance assessment remains constrained by fundamental evaluation limitations.

Methods rely on non-overlapping datasets with inconsistent protocols: many are proprietary, while public data undergo arbitrary resolution adjustments, confounding conditioning contributions with dataset-specific effects.
More critically, no standardized validation pipeline exists. These constraints prevent not only quantitative comparison within this survey but also reliable cross-work assessment in the literature itself—methods are rarely compared directly, obscuring which conditioning strategies, integration mechanisms, or architectural decompositions prove most effective and under what metrics. 
When numerical results are reported, metrics are applied non-uniformly: FID uses different feature extractors (ImageNet vs.\ RadImageNet), CLIP evaluations employ distinct encoders, and distributional and clinical assessments lack consensus on thresholds.
Conditioning claims are often accepted without systematic ablation studies, and computational costs remain disconnected from quality improvements, preventing principled trade-off analysis.

The field's rapid evolution (2023–2025) further complicates assessment: overlapping approaches—text-based versus report-based conditioning, ControlNet versus cross-attention integration—proliferate without principled comparison, obscuring incremental progress and leaving fundamental design questions unresolved. Despite these limitations, organizing literature into K$\times$I$\times$A 
design space reveals actionable opportunities and enables identification 
of concrete research directions, positioning future work toward cumulative 
refinement beyond continued proliferation of incomparable variants.

\section*{Ethical Statement}
There are no ethical issues.

\section*{Acknowledgments}
This paper is supported by the FAIR (Future Artificial Intelligence Research) project, funded by the NextGenerationEU program within the PNRR-PE-AI scheme (investment I.4.1) and by Fondazione Regionale per la Ricerca Biomedica (Regione Lombardia), project ID 012024R0055 PREDICT.

\bibliographystyle{named}
\bibliography{ijcai26}
\end{document}